\documentclass[aps,pra,twocolumn,showpacs]{revtex4}

\usepackage{physics,graphicx}
\newcommand{\gstate}{S$_{1/2}$ } 
\newcommand{\estate}{P$_{1/2}$ }
\newcommand{\dstate}{D$_{3/2}$ } 
\newcommand{\oa}{\omega_1 }
\newcommand{\ob}{\omega_2 }
\newcommand{\opp}{\omega_p }
\newcommand{\Ba}{Ba$^+$ }
\newcommand{\ipe}{ion-photon entanglement }
\begin{document}

\title{Ion-photon entanglement and quantum frequency conversion with trapped Ba$^+$ ions}

\author{J. D. Siverns, X. Li and Q. Quraishi*}
\affiliation{Army Research Laboratory, Adelphi, MD 20783 and \\ Joint Quantum Institute, University of Maryland, College Park, MD 20742}

*\email{*qudsia.quraishi.civ@mail.mil}

\date{compiled \today}

%\ociscodes{(270.5585) Quantum information and processing; (270.5565) Quantum communications; (020.0020) Atomic and molecular physics.}

%\doi{\url{http://dx.doi.org/10.1364/ao.XX.XXXXXX}}

\begin{abstract} 
Trapped ions are excellent candidates for quantum nodes, as they possess many desirable features of a network node including long-lifetimes, on-site processing capability and produce photonic flying qubits. However, unlike classical networks in which data may be transmitted in optical fibers and the range of communication readily extended with amplifiers, quantum systems often emit photons that have limited propagation range in optical fibers and, by virtue of the nature of a quantum state, cannot be noiselessly amplified. Here, we first describe a method to extract flying qubits from a \Ba trapped ion via shelving to a long lived, low-lying D-state with higher entanglement probabilities compared with current strong and weak excitation methods. We show a projected fidelity of $\approx$89\% of the ion-photon entanglement. We compare several methods of ion-photon entanglement generation and show how the fidelity and entanglement probability varies as a function of the photon collection optic's numerical aperture. We then outline an approach for quantum frequency conversion of the photons emitted by the \Ba ion to the telecom range for long-distance networking and to 780 nm, for potential entanglement with Rubidium based quantum memories. Our approach is significant for extending the range of quantum networks and for development of hybrid quantum networks compromised of different types of quantum memories.
\end{abstract}

%\setboolean{displaycopyright}{true}

\maketitle
%\thispagestyle{fancy}

%\ifthenelse{\boolean{shortarticle}}{\ifthenelse{\boolean{singlecolumn}}{\abscontentformatted}{\abscontent}}{}

\section{Introduction}
Trapped ions are a well-established system in which many of the desirable features for quantum information processing have been demonstrated \cite{Blatt13}, such as long-lived quantum coherence \cite{Szwer11,Langer05} and storage and retrieval of quantum states using photons on scalable platforms \cite{Kieplinski02}. From the first work \cite{Monroe96} demonstrating a quantum gate with one trapped ion to the engineering of quantum states with 14 ions \cite{Monz11}, efforts are now underway to establish networks of remotely situated trapped ion systems \cite{Monroe14}. Two quantum nodes are connected when entanglement has been established via a joint measurement of single photons (flying qubits), emitted by each quantum node separately \cite{Moehring07}. The nodes act as quantum memories, emitting photons entangled with the ion's internal state \cite{BBlinov04}. Entanglement between nodes can be established via Bell-state measurements. Such quantum networks may be used for teleportation between the nodes, even if there is no a-priori knowledge of the state to be teleported \cite{Olmschenk09}. Lab-based networks connected by approximately 1 km optical fibers have demonstrated entanglement but the flying qubit wavelenghts are incommensurate with low loss, long-distance propagation in optical fiber \cite{Hensen15}. Modern telecommunication (telecom) networks utilize well-established networks of optical fibers linking remotely situated nodes. Unlike classical approaches to extending these data ranges using optical amplifiers, quantum information can neither be cloned nor noiselessly amplified \cite{Wootters82}, however, two-node entanglement swapping can serve as a quantum repeater to transfer quantum information \cite{Briegel99}. 

It would be advantageous if emerging quantum networks could exploit existing telecom optical fiber infrastructures. In this case, long-distance quantum networking requires photonic flying qubits at telecom wavelengths for low-loss transmission. However, trapped ions, similar to other types of quantum memories like neutral atoms or nitrogen vacancy centers, typically emit photons which have high attenuation when propagated in optical fibers, as such memories emit either in in the ultraviolet (UV), visible or near-IR regime. Optical frequency conversion via nonlinear processes is well established and may be used to obtain more desirable wavelengths. Quantum frequency conversion (QFC)\cite{Kumar90} of single photons has been demonstrated using atomic vapors \cite{Radnaev10}, photonic crystal fibers \cite{McGuinness10} and crystals such as lithium niobate \cite{Rakher10,Pelc11}. Recent work has shown QFC into telecom wavelengths from single photons emitted by quantum memories such as quantum dots \cite{Zaske12} and neutral atoms \cite{Albrect14}. Importantly, the quantum nature of the photon is preserved in the quantum frequency conversion process \cite{Kumar90,Ates12,Albrect14}. In addition, creating entanglements between different types of memories in a hybrid quantum network allows for each system's unique properties to be utilized \cite{Waks09}. Here, we present an approach to extract polarization entangled photonic flying qubits from a $^{138}$\Ba trapped ion memory and outline a scheme for quantum frequency conversion for long-distance and hybrid quantum networking.

\begin{figure}%[htbp]
%\begin{center}
\includegraphics[width=1.0\columnwidth]{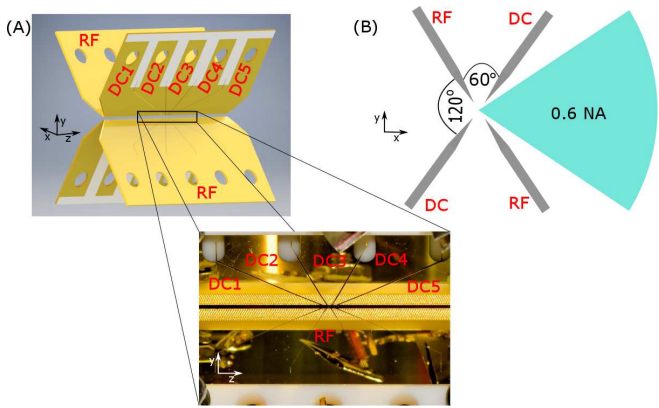}
\caption{\label{fig:trap}(a) Schematic of the blade trap. The distance between the upper (segmented into five static voltage electrodes) and lower (RF) blades is $\sim$260 $\mu$m and the width of the central electrode (labeled DC3) is $\sim$250 $\mu$m at its tip. For illustration purposes, the inset shows a close up photograph of the trap with more detail of the electrode segmentation. The segments of the rf blade they are electrically shorted together creating one electrode. (b) Diagram showing the trap geometry in the x-y plane with the light cone of the 0.6 NA lens (half-angle of around 37$^{\circ}$) shown in blue.}
\end{figure}

\section{Ion trapping and photon extraction}
\label{sec:ion}
Trapped ions confined in radio frequency (RF) Paul traps are versatile quantum systems from which flying qubits may be extracted with a high quantum correlation between the trapped ion memory and flying qubit \cite{BBlinov04}. Such systems have applications in frequency standards \cite{Bollinger91,Fisk97}, quantum information processing \cite{Kieplinski02,Debnath16,Haffner08}, quantum simulations \cite{JSmith16,Ivanov09,Porras04,Blatt12} and quantum networking \cite{LMDuan10,PMaunz09,AStute12}. Networking two-individual nodes has been implemented over a distance of meters \cite{LMDuan06,Moehring07,Hucul15}. Here, we propose a method for near deterministic ion-photon entanglement production using continuous wave (CW) excitation and compare this approach to methods using strong \cite{Maunz07} and weak excitation \cite{BBlinov04}. Strong excitation methods require a laser pulse with sufficient intensity to excite the ion with near unity probability. Typically, this is achieved with the use of a modelocked pulse laser \cite{Maunz07}. The approach outlined here has a higher entanglement probability compared with both strong and weak excitation schemes \cite{Auchter14,Slodicka13,BBlinov04}. The fidelity, although lower than other schemes, provides a higher entanglement probability which could be used for the demonstration of a proof-of-principle quantum network node without the need for additional experimental overhead involved with strong excitation schemes.

\subsection{Ion trapping background}
\label{sec:trapbasics}
RF voltages applied to trap electrodes provide a suitable ponderomotive pseudo-potential to trap ions of mass, $m$, and charge, $e$, which, for a linear Paul trap, is given by \cite{ghosh,Madsen04}

\begin{equation}\label{pond}
\psi=\frac{e^{2}V_{0}^{2}\eta^{2}}{4mr^{4}\Omega_{RF}^{2}}\left(x^{2}+y^{2}\right),
\end{equation}

\noindent
where $\Omega_{RF}/2\pi$ is the RF frequency, $r$ is the ion-electrode distance and $V_{0}$ is the amplitude of the RF voltage applied to the trap. Confinement of the ion along the z-axis can be achieved by the introduction of a potential in this direction created from various static voltage electrodes. The geometric factor $\eta$ \cite{Madsen04} is equal to one for a perfectly hyperbolic electrode geometry and less than one as the geometry strays from this perfect form. Ions trapped in this pseudo-potential will undergo secular motion with a frequency given by \cite{ghosh,Madsen04}

\begin{equation}\label{secular}
\omega_{s}=\frac{eV_{0}\eta}{\sqrt{2}mr^{2}\Omega_{RF}}.
\end{equation}

We use a linear four-blade trap as shown in Fig.~\ref{fig:trap}(a). Such traps possess a node in the ponderomotive potential along the axis of the trap allowing either a single ion or a chain of ions to be trapped. Confinement along this node is provided by applying static voltages to the outer segments of the electrodes. These traps typically have depths of 1 eV to 10 eV, corresponding to $\approx$10$^5$K. It is possible to achieve ion temperatures of approximately 1 mK by applying a single laser cooling beam with momentum projections in all three motional directions of the ion. Hence, the ion can remain tightly confined in the trap allowing for repeated interrogation by optical beams for single photon production.

To increase the ion-photon entanglement probability, collection optics must capture as much spontaneously emitted light as possible. We have custom-designed our apparatus with a particularly high collection aperture optic. The physical arrangement of the blades allows for collection up to $\approx$10$\%$ (numerical aperture (NA) of 0.6) of the light emitted from ions in the trap as shown in Fig.~\ref{fig:trap}(b). Similar lenses integrated with blade-traps have substantially improved ion-photon entanglement probabilities \cite{Hucul15}. 

\subsection{Ion state preparation and ion-photon entanglement}
\label{sec:photonextract}
\begin{figure}[htbp]
\includegraphics[width=\linewidth]{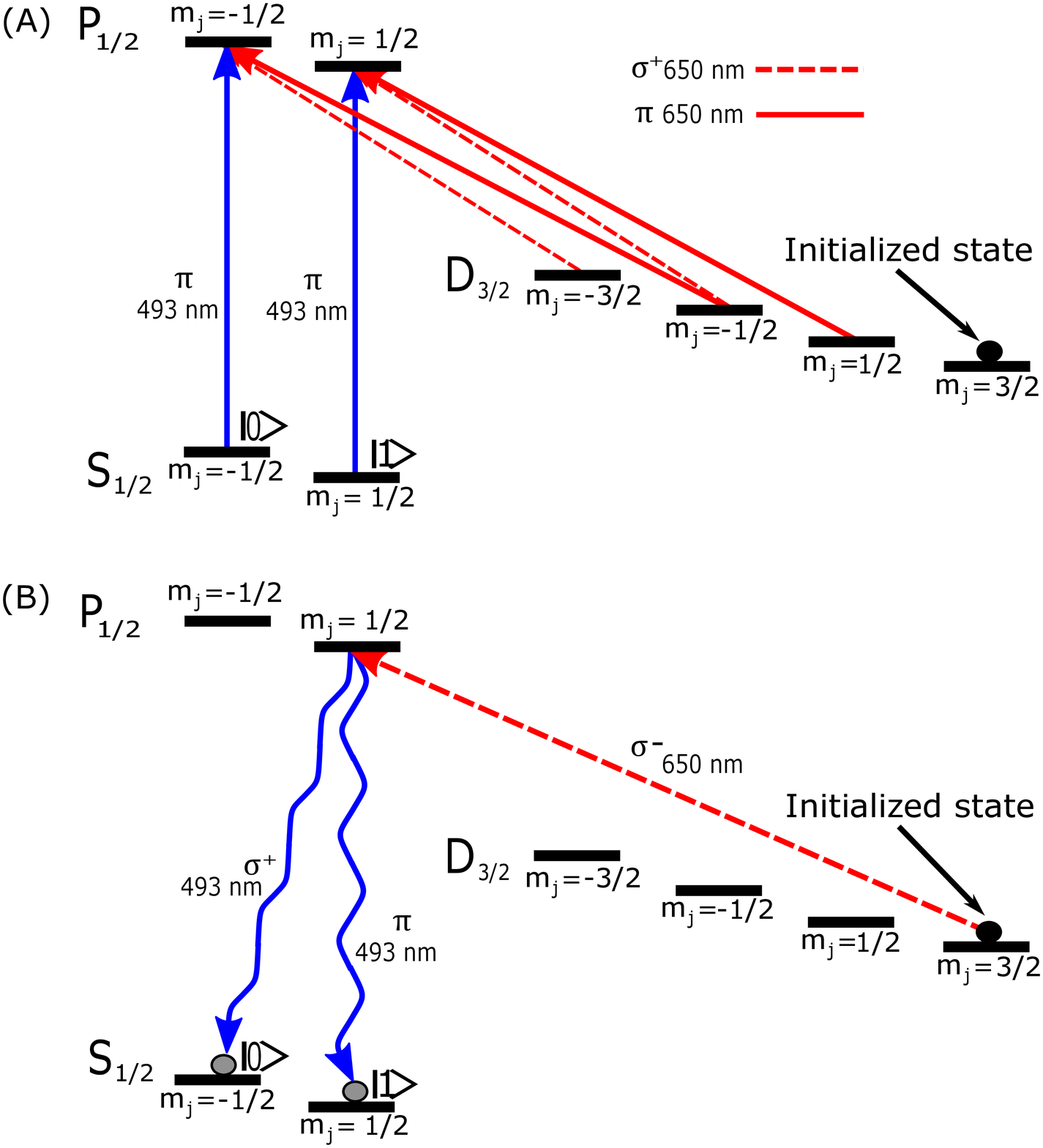}
\caption{(a) The transitions required for initializing a $^{138}$Ba$^+$ ion in the $m_{j}=+3/2$ state in the \dstate level. (b) The  650 nm $\sigma_-$ transition required for the generation of the 493 nm photon and production of the ion-photon entangled state given in equation \ref{ip_entanglement}.}
\label{fig:levels}
\end{figure}

A qubit is represented by the two Zeeman levels in the $S_{1/2}$ ground state of a $^{138}$Ba$^+$ ion with $\ket{0}=\ket{m_{j}=-1/2}$ and $\ket{1}=\ket{m_{j}=+1/2}$ as shown in Fig.~\ref{fig:levels}(a). The ground state, \gstate, is split into two Zeeman levels by applying a magnetic field which defines the direction of the quantization axis. It is possible, as we show below, to produce a single 493 nm photon with its polarization entangled with the qubit states $\ket{1}$ and $\ket{0}$ via shelving in the \dstate level. The \dstate level is long-lived (lifetime of 80 sec) allowing for effective shelving. 

The \ipe is produced using the following three steps. Firstly, the ion is initialized into the $m_{j}=+3/2$ ($m_{j}=-3/2$) state in the \dstate level, using $\pi$-polarized 493 light along with $\pi$ and $\sigma^+$ ($\sigma^-$) polarized 650 nm CW lasers. Secondly, the ion is excited from the \dstate state to the \estate state via a $\sigma^-$ ($\sigma^+$) polarized 650 nm beam. Initially, the ion is only excited to the \estate state Zeeman level $m_j=+1/2$ ($m_j=-1/2$). The effects of multiple excitations, including those to the opposite Zeeman level will be discussed in section \ref{sec:multiexcite}. Finally, the ion can spontaneously decay from the excited state, \estate, to the ground state, \gstate, via two dipole-transition allowed paths corresponding to either a $\pi$ or $\sigma^+$ polarized photon. Observation in the direction perpendicular to the quantization axis means it is possible to view $\sigma^{\pm}$ and $\pi$ photons as horizontal $\ket{H}$ and vertical $\ket{V}$ polarizations respectively, yielding an ion-photon entangled state given by

\begin{equation}
\ket{\Psi_G}=\frac{1}{\sqrt{2}}\ket{V}\ket{1}+\frac{1}{\sqrt{2}}\ket{H}\ket{0}.
\label{ip_entanglement}
\end{equation}

The state given by Eqn.~\ref{ip_entanglement} takes into account both the intensity pattern of the emitted $\sigma^+$ and $\pi$ polarizations, with the latter being greater by a factor of two when the spherical polar angle, $\theta$, of the emitted photons, with respect to the dipole axis, is equal to $\pi/2$. Also included are the Clebsch-Gordan coefficients of the transitions, shown in Fig.~\ref{fig:dtop}(a). 

This intensity pattern difference arises from our observation direction with respect to the quantization axis. Photons emitted from the \estate to \gstate transition have the unnormalised polarization states, $\ket{\pi}$ and $\ket{\sigma^{\pm}}$, which are given by

\begin{equation}\label{sigmapol}
\ket{\pi}=-\sin\theta\ket{\hat{\theta}}
\end{equation}

\noindent
and 

\begin{equation}\label{pipol}
\ket{\sigma^{\pm}}=\frac{e^{\pm i\phi}}{\sqrt{2}}\left(\cos\theta\ket{\hat{\theta}}\pm i\ket{\hat{\phi}}\right)
\end{equation}

\noindent
respectively, where $\phi$ is the azimuthal angle with respect to the dipole axis and $\hat{\theta}$ and $\hat{\phi}$ are the respective unit vectors. It can be seen from Eqns. \ref{sigmapol} and \ref{pipol} that when the angle $\theta$ is set to $\pi/2$ the two polarisations are orthogonal \cite{BBlinov04} and, after taking into account the relevant Clebsch-Gordan coefficients, we obtain Eqn.\ref{ip_entanglement}. 

It should be noted that initialization into the opposite Zeeman level of the \dstate ($m_j=-3/2$) is driven by $\sigma^+$ polarization of the 650 nm beam. This results in coupling to the  $m_j=-1/2$ Zeeman level in the \estate level and produces an ion-photon entangled state given by

\begin{equation}\label{psibad}
\ket{\Psi_B}=\frac{1}{\sqrt{2}}\ket{H}\ket{1}+\frac{1}{\sqrt{2}}\ket{V}\ket{0}.
\end{equation}

The flying qubit in this state carries with it information about the atom's state, meaning that a measurement of one polarization correlates to a particular atomic state. Interference between photons from a similarly prepared remote quantum memory node would generate entanglement between the two nodes, forming a quantum network. Two prominent factors which would degrade the purity of the ion-photon entangled state include re-excitation from the \dstate level and polarization mixing, as discussed in sections \ref{sec:multiexcite} and \ref{sec:photoncollect} respectively.

\subsection{Fidelity of ion-photon entanglement}\label{fidelity}
The target state given by Eqn.~\ref{ip_entanglement}, is produced by decay from the $m_j=1/2$ state of the P$_{1/2}$ level to the two S$_{1/2}$ ground states via two paths with Clebsch-Gordan coefficients shown in Fig.~\ref{fig:dtop}(a) (solid blue line). Factors which degrade the purity of the ion-photon entanglement include re-excitation from the \dstate level and polarization mixing due to the large collection angle. In the former case, rather than a pure state, a mixed state is obtained affecting the ability of the scheme to generate the desired entanglement. A common measure of the extent to which the desired target state is obtained is given by the fidelity. Below we analyze the fidelity under re-excitation and polarization mixing. 

\subsubsection{Effect of multiple excitations on state fidelity}\label{sec:multiexcite}

\begin{figure}[htbp]
\includegraphics[width=\linewidth]{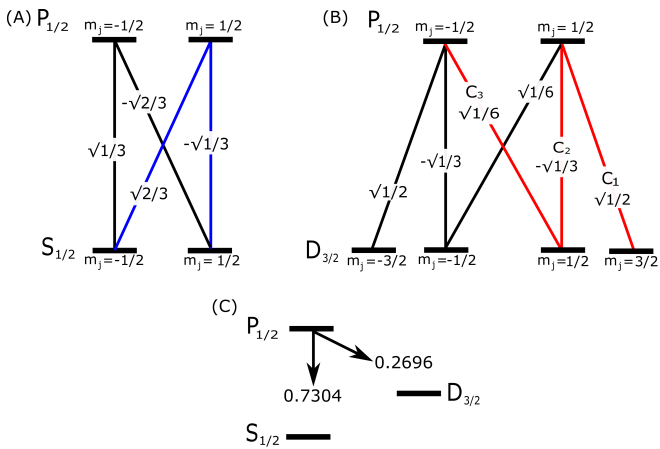}
\caption{Diagram showing the Clebsch-Gordan coefficients for transitions between the (a) ground-state S$_{1/2}$ and \estate levels and (b) the \estate and meta-stable \dstate level. (c) A simplified energy level diagram showing the branching ratio from the excited \estate level to the S$_{1/2}$ and the \dstate level \cite{Dmunshi15}.}
\label{fig:dtop}
\end{figure}

The excitation scheme presented here (Fig.~\ref{fig:levels}) is unlike some \cite{BBlinov04,Auchter14} used for ion-photon entanglement as the excitation (at 650 nm) is at an entirely different wavelength than the emitted photon (at 493 nm). This is clearly advantageous in terms of filtering background light, but also serves to improve ion-photon entanglement probabilities. Re-attempts at successful entanglement can occur after failed attempts during the same experiment cycle, albeit at the expense of the fidelity. 

The ratio of a CW pulse duration, $\Delta t$, to the natural linewidth, $\tau$, gives the probability of double excitation, $p_{\gamma} = 1-\exp(-\Delta t/\tau)$. If we assume a worst case scenario of continuous ($\Delta t\rightarrow\infty$) exposure  to 650 nm $\sigma^-$ light, then $p_{\gamma}\rightarrow 1$. The re-excitation occurs because there is an appreciable probability (approximately 27\%, see Fig.~\ref{fig:dtop}) of the ion decaying back to the \dstate manifold, although not exclusively into the initialized state shown in Fig.~\ref{fig:levels}(a). The decay back to the \dstate manifold happens via three possible channels emitting either a $\sigma^+$, $\sigma^-$ or $\pi$ photon, with the Clebsch-Gordan (C-G) coefficients shown in Fig.~\ref{fig:dtop} \cite{Dmunshi15}.   

If the ion decays back to the initialized state shown in Fig.~\ref{fig:levels}(a), we would proceed as already described in Section \ref{sec:photonextract} without any effect on the fidelity. If the decay is via a $\pi$ photon (to the $m_{j}=+1/2$ Zeeman level, with C-G coefficient of $-\sqrt{1/3}$) then the subsequent 650 nm excitation will be to the $m_{j}=-1/2$ Zeeman level of the \estate state. Any 493 nm decay to the ground state will result in the ion-photon entangled state given by Eqn.~\ref{psibad} being created instead of the desired state given by Eqn.~\ref{ip_entanglement}. If the decay is via a $\sigma^+$ photon (to the $m_{j}=-1/2$ level, with C-G coefficient $\sqrt{1/6}$), the ion will go dark and no ion-photon entanglement will occur. Therefore, the ion-photon entanglement created with this scheme will be given by the mixed state

\begin{equation}\label{ionphoton}
\Psi_{m}=\frac{P_G}{P_G+P_B}\ket{\Psi_G}+\frac{P_B}{P_G+P_B}\ket{\Psi_B}
\end{equation}

\noindent
where $P_G$ and $P_B$ are the probabilities of the scheme creating the ion-photon entangled states $\ket{\Psi_G}$ and $\ket{\Psi_B}$ respectively, and are given by the following geometric series

\begin{equation}\label{Pg}
P_G=\frac{Br_{493}}{1-(C_1^2Br_{650})}\approx 0.844
\end{equation}
\begin{equation}\label{Pb}
P_B=\frac{Br_{493}Br_{650}C_2^2}{1-(C_3^2Br_{650})} \approx 0.103.
\end{equation}

\begin{figure}[htbp]
\centering
\includegraphics[width=\linewidth]{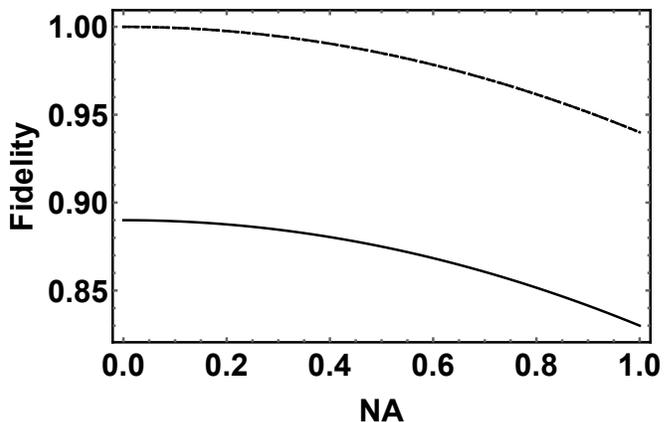}
\caption{The fidelity of the ion-photon entangled state as a function of NA is shown for the scheme outlined in Section \ref{sec:photonextract} (solid), weak and strong (both dashed) excitation methods. The weak and strong excitation methods fall on the same upper (dashed) curve.}
\label{fig:fidNA}
\end{figure}

Here $Br_{493}$ and $Br_{650}$ are the branching ratios from the \estate state to the ground state and \dstate state respectively, and $C_1$, $C_2$ and $C_3$ are the C-G coefficients of the decays shown in Fig.~\ref{fig:dtop}. This means that a 493 nm photon will be emitted by the ion with a success of $\approx 94.7\%$, making the scheme near deterministic, albeit with a lower fidelity for the target state when compared with weak and strong excitation methods as shown in Fig. \ref{fig:fidNA}. When unsuccessful, the ion will have been pumped into a dark state in the \dstate level. 

Here, we examine the effect of these re-excitations by using the metric of fidelity, defined as \cite{Jozsa94} 

\begin{equation}\label{fid}
F=\bra{\Psi}\rho\ket{\Psi},
\end{equation}

\noindent
where $\ket{\Psi}$ is the intended pure state and $\rho$ is the density matrix of the created state. The fidelity of the mixed state, $\Psi_{m}$, created by this scheme with the intended pure state, $\Psi_G$, can be calculated to be $\approx 0.89$. Although this scheme yields a lower fidelity than standard strong and weak excitation methods (both can achieve fidelities near to unity), it allows the user to produce ion-photon entanglement with a higher probability and in the case of the strong excitation method, without the need for additional pulsed laser apparatus. By adopting this method we will have the ability to perform proof-of-principle quantum networking experiments with quantum frequency conversion of the emitted photon. 

\subsubsection{Effect of polarization mixing due to collection solid-angle on state fidelity}
\label{sec:photoncollect}
Ion-photon entanglement is useful as a resource for quantum networking, as the photon carries information about the atomic state to a remote site. To optimize networking protocols we wish to capture photons from as many entanglement trials as possible. However, the photon is emitted from the \estate level spontaneously into the full solid angle 4$\pi$. There are a variety of approaches to optimize collection including optical cavities \cite{Casabone15,Begley16} and in-vacu \cite{Streed11,Shu09} and ex-vacu \cite{Hucul15} high NA lenses. High NA lenses can have such a large collection angle that the $\sigma^{\pm}$ polarized light typically has a projection on the $\pi$ polarized light when viewed in a horizontal and vertical polarization basis. Here we examine how this projection reduces the fidelity of the \ipe\!. 

We plan to integrate an ex-vacu 0.6 NA lens into our setup which is designed to collect approximately 10\% of the light and is AR coated for both 493 nm and 650 nm. Ex-vacu high NA collection reduces complexity of the in-vacuum trap assembly and readily allows for optical corrections with additional free-space optics. Although, in-vacuo optics can provide a more modular, compact and scalable solution for future nodes of a potential quantum network \cite{Mehta16,Streed11}. The analysis below holds for both types of lenses. 

The fraction of emitted photons collected by a lens with numerical aperture, NA, is given by

\begin{equation}\label{NA}
\frac{\Omega}{4\pi}\approx\frac{1}{4}\mbox{NA}^2.
\end{equation}

\noindent
From Eqn.~\ref{NA} it is clear to see that a lens with a high NA will result in the collection of a higher fraction of the photons emitted from an atom. However, as can be seen in Eqns.~\ref{sigmapol} and \ref{pipol}, if the spherical polar angle, $\theta$, is not equal to $\pi/2$ then there will be a loss of orthogonality between the $\sigma$ and $\pi$ light emitted from the atom, when viewed in a Cartesian basis. In this case, the fidelity of the collected ion-photon entangled state varies from the ideal ion-photon entangled state. For CW excitation of the ion as outlined in \ref{sec:photonextract}, the fidelity of the entangled state as a function of the collection lens's NA (Fig.~\ref{fig:fidNA}, solid trace) is given by \cite{BBlinov04}

\begin{equation}\label{fidNA}
F=F_{\mbox{max}}-0.24(\Omega/4\pi)
\approx F_{\mbox{max}}-0.24\left(\frac{1}{4}\mbox{NA}^2\right)
\end{equation}

\noindent
where $F_{\mbox{max}}$ is the maximum fidelity achievable. For the method outlined in section \ref{sec:photonextract}, the fidelity $F_{\mbox{max}}\approx$ 0.89, so we obtain a fidelity $F_{0.6}=0.87$ for NA $=0.6$. For comparison, also shown are curves for a weak excitation method \cite{Auchter14} and a strong excitation method \cite{Maunz07} (dashed line). Both have the same fidelity as both have $F_{\mbox{max}}$ near unity.

The approach presented here does not reach the high fidelities (unity with a NA of zero) achieved with these other two schemes, however, it does yield a higher probability of entanglement compared to the other two schemes as shown below.
For improved fidelities, achieved by avoiding multiple excitations, using a single pulse from a pulsed laser is a desirable approach even though the entanglement probability is reduced.

\subsubsection{Probability of \ipe with \dstate initialization}
\begin{figure}[htbp]
\centering
\includegraphics[width=\linewidth]{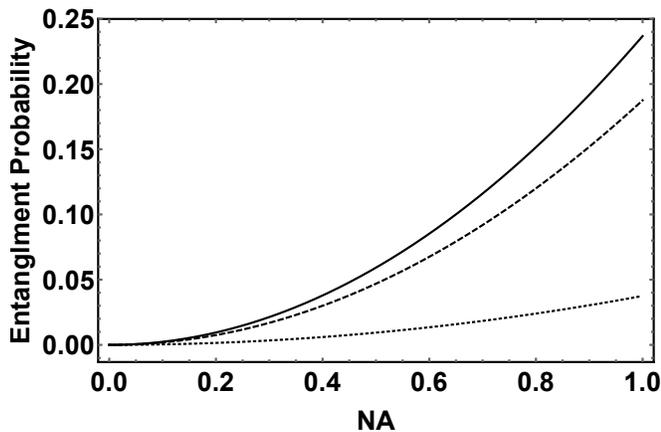}
\caption{The entanglement probability of the scheme outlined in Section \ref{sec:photonextract} (solid), weak excitation method (dotted) and strong excitation method (dashed) as a function of NA.}
\label{fig:probNA}
\end{figure}

Using a high NA lens increases the probability of photon capture and hence, the \ipe probability. This higher entanglement probability is shown in Fig.~\ref{fig:probNA} and is given by

\begin{equation}\label{probNA}
P=P_eP_s\frac{\mbox{NA}^2}{4}
\end{equation}

\noindent
where $P_e$ is the probability of excitation to the P-level and $P_s$ is the probability of decay to the S-level. For the curves shown, the weak and strong excitation methods have $P_e$ equal to 0.2 and 1 respectively and both have $P_s$ equal to 0.7304, set by the branching ratio [Fig.~\ref{fig:dtop}(c)]. For the method outlined in Section \ref{sec:photonextract} where we initialize in the \dstate\! level, the product, $P_eP_s = 0.947$ (as shown in Section \ref{sec:multiexcite}) meaning re-excitation increases the probability of obtaining an entangled state. For an NA $=0.6$  an \ipe probability of  $P=0.085$ can be obtained. It is notable that the excitation scheme outlined in Section B has a reasonable fidelity but the highest entanglement probability. For proof-of-principle experiments this would be appropriate to transmit quantum information to a remote site.

\begin{table}[htbp]
\centering
\caption{Comparison of excitation schemes with fidelity, $F_{0.6}$, and ion-photon entanglement generation probability, $P_{0.6}$, evaluated at NA$=0.6$.}
\begin{tabular}{cccc}
\hline
Excitation scheme & $P_ePs$ & $P_{0.6}$ & Fidelity,$F_{0.6}$ \\ 
\hline
  \dstate excitation & 0.947 & 0.085  & 0.87 \\
  Weak excitation & 0.146 & 0.014 & 0.98\\
  Strong excitation & 0.730 & 0.068 & 0.98\\
\hline
\end{tabular}
\label{table}
\end{table}

\section{Quantum frequency conversion of single photons}
\begin{figure}%[fiberattn]
	\centering
	\includegraphics[width=\linewidth]{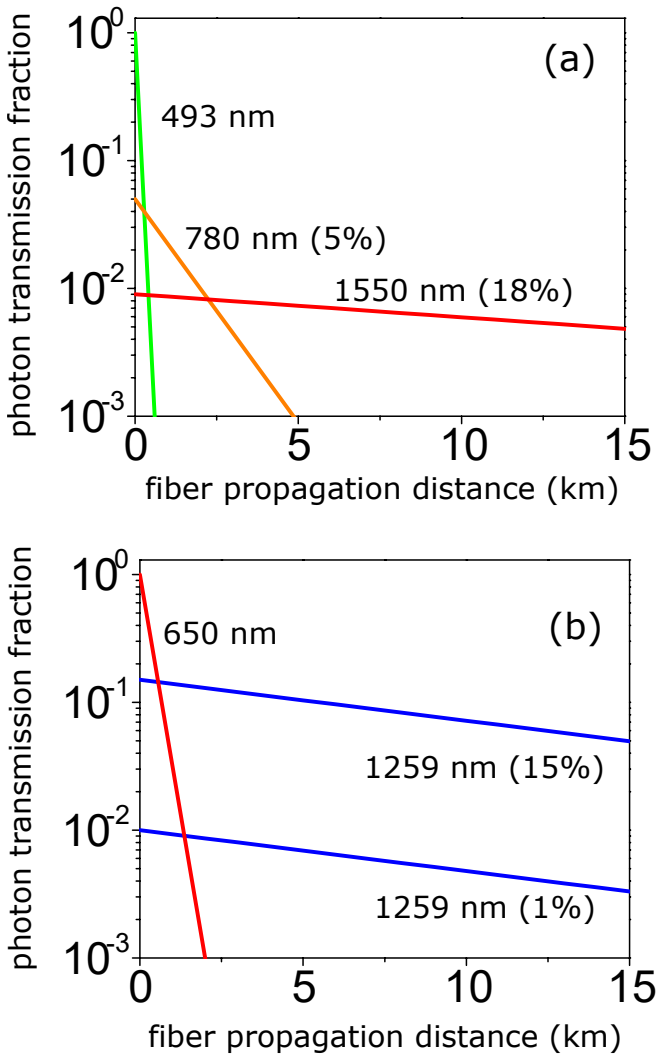}
	\caption{Photon transmission fraction as a function of distance though wavelength specific fiber of photons derived from \Ba ions. Representative quantum frequency conversion efficiencies are given in parenthesis. (a) (green trace) Transmission of 493 nm photons with a fiber loss of -50 dB/km. (orange trace) Transmission of 780 nm photons with a fiber loss -3.5 dB/km, from frequency converted 493 nm photons. (red trace) Transmission of 1550 nm photons with a fiber loss -0.18 dB/km, from frequency converted 780 nm photons ($\sim$18\% QFC value reported in \cite{Albrect14})  (b) (red trace) Transmission of 650 nm photons (fiber loss of -15 dB/km). (blue trace) Transmission of 1259 nm photons with a fiber loss of -0.3 dB/km from 650 nm photons.}
	\label{fiberattn}
\end{figure}

Trapped ions can serve as excellent quantum memories, however, they often emit in optical frequency ranges that have substantial attenuation even in dedicated wavelength-specific optical fibers. For instance, Yb$^+$ ions are highly desirable qubits, but they emit photons at 369 nm which have approximately 70 dB/km attenuation in commercially available single mode fibers. Although Ba$^+$ wavelengths (493nm, 650 nm) have improved fiber transmission over this Yb$^+$ emission wavelength, there is still substantial attenuation, as shown in Fig.~\ref{fiberattn}.  

Quantum frequency conversion \cite{Kumar90} may be implemented to convert photons from the ion into the telcomm regime for low loss propagation through optical fibers therefore extending the range of quantum networking from lab-based two-node setups \cite{Maunz07,Bernein13}. Additionally, spectrally mismatched photons may be frequency converted to establish a hybrid quantum network comprised of two different types of quantum memories \cite{PhysRevLett.109.147405}. The frequency conversion must preserve the quantum properties and indeed, QFC experiments of single photons have shown that the quantum state of the single photon was preserved after QFC \cite{Rakher10,McGuinness10}.

Here, we present an approach for QFC of \Ba ion photons emitted at 493 nm and 650 nm. The 493 nm photons will be converted into the near infra-red (NIR) regime for potential networking with a neutral atom based quantum memory and the 650 nm photons will be converted into the telecom regime for long-distance quantum information transmission. 

\subsection{Quantum frequency conversion (QFC) background}
Quantum frequency conversion \cite{Kumar90} is a nonlinear process and relies on parametric oscillations of a material's electronic susceptibility, $\chi$. With sufficiently intense optical pump fields, the higher order components of $\chi$ begin to dominate and modulate the material's polarizibility as a function of the applied field, given by

\begin{equation}\label{eqn:polarise}
P_i =  \varepsilon_0\sum\limits_{j}\chi_{ij}^{(1)} E_j+\varepsilon_0\sum\limits_{jk}\chi_{ijk}^{(2)}E_jE_k + \varepsilon_0\sum\limits_{jkl}\chi_{ijkl}^{(3)}E_jE_kE_l+\cdots
\end{equation}

\noindent
where $ \varepsilon_0 $ is the free space electric permittivity, $E_{ijkl}$ denotes the electric field, and $ \chi^{n}$ is a rank $n+1$ tensor.

In a media with a large $\chi^{(2)}$ coefficient, the second order term in Eqn.~\ref{eqn:polarise} can dominate with a strong pump field, allowing the non-linear three-wave mixing (TWM) process to occur. Here the generation of the frequency $ \ob = \oa  \pm \opp$ is expected, where $\oa$ is the frequency of the light to be converted and $\opp$ is the frequency of the high intensity pump. The plus sign corresponds to sum frequency generation (SFG), or upconversion, while the minus sign means difference frequency generation (DFG), or downconversion. Examples of a TWM process include frequency doubling in lithium niobate (LN), beta-barium-borate (BBO) and lithium triborate (LBO) crystals where $\opp=\oa$ and $ \ob = 2 \oa $. With quasi-phase matching (QPM), TWM can be achieved in periodically poled materials like lithium niobate and potassium titanyl phosphate (KTP) even when $\opp\neq\oa$.

It is also possible to perform four-wave mixing (FWM) in centrosymmetric media as the second order term in Eqn.~\ref{eqn:polarise} is zero due to symmetry and the third order $\chi^{(3)}$ term can dominate, when subject to two pump lasers of sufficient intensity. $\chi^{(3)}$ media such as silicon nitride (SiN) microresonators \cite{Vernon16} and optical fibers can both support FWM \cite{McKinstrie05}. Recent work in SiN \cite{Qing16} has shown conversion efficiencies on par with those observed in PPLNs, however, not at a low enough noise rate required for the relatively low photon rates emitted by atomic quantum memories.

Frequency conversion at the single photon level with sufficiently low dark counts has been shown using TWM in periodically poled waveguides \cite{Albrect14}.  Given expected photon data rates \cite{Hucul15}, we plan to use a TWM process for frequency conversion of the \Ba ion photons.  

In the case of periodically poled materials suitable for TWM, the quasi-phase matching can be achieved when

\begin{equation}
k_1-k_p-k_2 - 2\pi m/\Lambda = 0 
\label{eqn:phase-matching}
\end{equation}

\noindent
where $ k_1 = \oa n_1/c$, $k_2 = \ob n_2/c$, $k_p = \opp n_p/c$, $m$ is the poling order and $\Lambda$ is the poling period. QPM is achieved by periodically-poling the material and yields more than an order of magnitude improved conversion efficiencies as compared against conversion using the largest coefficient that can be phase-matched with birefringence \cite{Boyd08}. 

Most work in quantum frequency conversion has been done using TWM in periodically-poled lithium niobate (PPLN) because it can be readily engineered to yield high conversion efficiencies over a range of input wavelengths, especially with waveguides fabricated in them. Using a reverse proton exchange (RPE) technique \cite{Parameswaran02} a waveguide may be buried beneath the surface of the bulk crystal to reduce transmission loss and to ensure better spatial mode confinement of all the optical beams within the PPLN. Another desirable feature of these crystals are the observed low noise counts, suitable for the signal level emitted by atomic quantum memories \cite{Albrect14}. 

Given that we are interested in converting and detecting single photons, noise photons at $\oa$ and $\ob $ must be minimized as they can spoil the \ipe\!, considering current entanglement probabilities \cite{Hucul15}. The primary sources of noise are spontaneous Raman scattering (SRS) and spontaneous parametric downconversion (SPDC). The SRS scattering can produce Raman peaks of noise sitting atop a noise pedestal. These effects are significantly minimized by ensuring $\ob $ is greater than $\opp $ and by sufficient $\ob-\opp $ detuning \cite{Pelc11}. The SPDC noise may be minimized by ensuring that the pump wavelength is the longest wavelength in the TWM process \cite{Pelc11} so that parasitic photons are produced at lower frequencies than either the input $\oa $ or converted signal $\ob $.  A final step in minimizing noise from the pump is done using optical filters, along with optical prisms and fiber Bragg gratings \cite{Albrect14}, to ensure that the background is significantly lower than the converted signal. Conversion of coherent light at 369 nm has been done to telecom wavelengths with a single-stage DFG process, but with $\opp$ greater than $\ob$ \cite{Rutz2016}, resulting in SPDC noise at the target telecom photon, and also in a two-stage process \cite{6000036} with higher poling period and, therefore, reduced conversion efficiencies \cite{Clark12}. To observe a long-lived memory entangled with either a visible or telecom photon, it should be possible to generate local entanglement between Yb$^+$ and Ba$^+$ \cite{Wright2016} and then do QFC on the photon emitted by the Ba$^+$ ion.  

\subsection{Proposed quantum frequency conversion for long distance and hybrid networking with \Ba ions}
\Ba has two strong dipole transitions (493 nm, 650 nm) from the \estate state. We begin by discussing frequency conversion of the 493 nm photons. In a single DFG stage, the 493 nm photon may be combined with a strong pump beam at 1343 nm ($\opp/2\pi =$ 223 THz) to produce a photon at 780 nm ($\ob/2\pi = $ 384 THz). This target wavelength is chosen to match the 780 nm D2 transition in neutral $^{87}$Rb atoms.  The pump laser may be frequency stabilized to match the 780 nm photon to the $^{87}$Rb D2 transition. The similarity of the temporal profiles of the photons emitted from \Ba and Rb-based quantum memories, means that entanglement between them might be created via Bell state measurements \cite{PhysRevLett.91.110405}, therefore creating a hybrid ion-neutral two-node quantum network that utilizes properties of each memory.

For long-distance communication through optical fibers, we can envisage using a second DFG conversion stage to convert the 780 nm photon to 1551 nm, with a pump at 1569 nm. Indeed, 780 nm photons from Rb-based quantum memories have already been converted into the telecom regime \cite{Albrect14,Radnaev2010}. In Fig.~\ref{fiberattn}(a) we show the propagation losses through wavelength specific fiber given conservative estimates on the QFC efficiencies. Given the high propagation loss of 493 nm in optical fiber, it is advantageous to do a single stage conversion efficiency to 780 nm for propagation over approximately 0.5 km even with a modest 5\% QFC efficiency.

A \Ba ion's 650 nm ($\oa/2\pi =$ 461 THz) photon can be converted into the telecom regime in a single DFG stage with the same CW pump laser at 1343 nm. The converted photon will be at 1259 nm ($\ob/2\pi = $238 THz), near the beginning of the telecom o-band. Although there is a smaller probability of the ion emitting a 650 nm photon over a 493 nm photon (see Fig.~\ref{fig:dtop}), we only need a single DFG stage to obtain a telecom photon. However, two-stage conversion is also possible as shown in recent work of conversion from 650 nm to 1550 nm \cite{Esfandyarpour:16}. In Fig.\ref{fiberattn}(b) we show the propagation losses through wavelength specific fiber of 650 nm and the target frequency converted photon given a modest conversion efficiency. Two different values of QFC efficiency are plotted to illustrate the crossing points between $\oa$ and $\ob$ at which conversion is beneficial. The approach outlined here for single-stage conversion of each \Ba ion allows for hybrid and long-distance quantum communication. 

In Table \ref{tab:qfc} we summarize the relevant potential conversion possibilities. Although PPLN crystals have typically shown the best QFC performance, we will use PPKTP for the 493 nm conversion as it has a higher photorefractive damage threshold. The damage threshold is not relevant for single photon input to the PPKTP but is relevant for the coherent light levels needed for initial alignment.
 
\begin{table}[htbp]
\centering
\caption{\bf QFC materials and frequencies}
\begin{tabular}{ccccc}
\hline
Conversion & $\oa/2\pi$ & $\ob/2\pi$ & $\opp/2\pi$  & Device \\
    & (THz)& (THz)& (THz)& \\
		\hline
493 nm $ \rightarrow$ 780 nm & 608  & 384  & 223  & PPKTP\\
650 nm $ \rightarrow$ 1259 nm  & 461 & 238  & 223  & PPLN \\
780 nm $ \rightarrow$ 1550 nm& 384  & 193 & 191 & PPLN \\
\hline
\end{tabular}
  \label{tab:qfc}
\end{table}

\subsection{Planned experimental approach for visible photon frequency conversion}
Previously we established a frequency converter setup for a neutral atom wavelength \cite{Li16}. We can extend this approach to the \Ba wavelengths of interest here. To observe frequency conversion (and hence obtain the QFC conversion efficiency), we plan to use CW tunable external cavity diode laser (ECDL) sources. One ECDL is at the input frequency $\oa$ (493 nm or 650 nm) and the other at the pump $\opp$ (1343 nm), seeding a single frequency Raman fiber amplifier. The seed laser can be locked to a transfer cavity, so that the converted photon is on resonance with the Rb transition.

\begin{figure}[htbp]
	\centering
	\includegraphics[width=\linewidth]{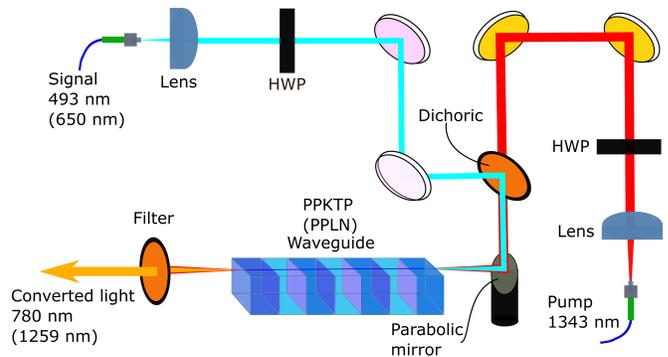}
	\caption{Diagram showing the proposed DFG setup. The signal and pump beams are collimated out of each fiber with an aspheric lens. Their polarization are adjusted with half-wave-plates (HWP) to vertical, i.e., perpendicular to the PPKTP chip surface with waveguides. The two beams are combined on a dichroic mirror before being simultaneously focused into the waveguide by a parabolic mirror. Residual pump and signal are filtered to allow measurement of the converted photons.}
	\label{fig:DFGSetup}
\end{figure}

For efficient coupling of both $\oa$ and $\opp$ into the nonlinear waveguide, we plan to implement a free space scheme with a silver-coated, off-axis, parabolic mirror with a 15 mm focal length, as shown in Fig.~\ref{fig:DFGSetup}. This gives us the ability to simultaneously couple the visible light and 1343 nm light into the respective fundamental modes of same waveguide \cite{Li16}. Parabolic mirrors are excellent for focusing beams of vastly different wavelengths to the same spot, as they posses no chromatic aberration. Before the parabolic mirror, the two collimated beams are combined on a dichroic mirror that reflects signals at 493 nm or 650 nm and passes the 1343 nm pump. The waists of the beams are chosen to match the mode field diameter (MFD) of the fundamental mode of the relative color inside the waveguide.

To align the beams into the waveguide, we first put the two colors into a wavelength-division multiplexing (WDM) fiber combiner and put the output near the focus of the parabolic mirror. The output is then back-propagated through the setup. We adjust WDM along with the parabolic mirror on a 3-axis translation stage to collimate both colors simultaneously. We then separate the two colors with a dichroic mirror, and couple each beam into their relative fiber. As a result, if we send each color back through the fibers, they will be focused at the same spot after the parabolic mirror. The WDM is then exchanged with a nonlinear chip in a temperature controlled oven. By only adjusting the nonlinear chip's position with a 5-axis translation stage, we can couple both beams into the same waveguide.

There are various free space coupling schemes that use other focusing optics, e.g., aspheric lens, grin lens, etc. However, in order to simultaneously couple both colors well, one either has to pre-shape one of the beams or to custom design the lens to be diffraction limited for both $\oa$ and $\opp$. It is also possible to write WDM waveguides before the PPKTP/PPLN waveguide. This way, optical fibers carrying each color can be directly attached to the input of the corresponding WDM waveguide. Such miniaturized setups have excellent coupling efficiencies and are highly scalable. However, they are wavelength specific and typically not commercially available. 

For QFC of single photons from the \Ba ion, the background noise from the pump laser needs to be carefully filtered. Narrow bandpass filters along with optical prisms and fiber Bragg gratings \cite{Albrect14} can be used to ensure that the background is significantly lower than the converted signal. The final single photon detection can be performed with a superconducting nanowire single photon detector, which can achieve more than 95\% quantum efficiency in the telecom regime \cite{Lita:08}.

\subsection{Summary}
Establishing a long-distance network using quantum memories involves entanglement generation between nodes. Photons entangled with quantum memories are excellent carriers of quantum information. Our approach to extract flying qubits from a \Ba ion provides high entanglement probabilities between the photon and ion compared with current weak and strong excitation schemes, however, at the loss of some fidelity. We showed that although high numerical aperture lenses can improve the photon collection efficiency, they can act to degrade the quality of the \ipe state due to polarization mixing. We proposed quantum frequency conversions of the \Ba ion wavelengths with only one pump laser to drive both conversion processes. The method outlined would produce photons available either for hybrid quantum networking or long-distance quantum communication utilizing existing telecom fiber networks.

\section{Funding Information}
Funding provided by the Army Research Laboratory (ARL) under Cooperative Agreement (W911NF-14-2-0101) and ARL's Center for Distributed Quantum Information. 

\section{Acknowledgements}
We thank Chris Monroe for the use of the ion trap blades shown in our trap photograph and we thank Martin Lichtman for a thorough reading of the manuscript.

\bigskip
\noindent
\bibliographystyle{unsrt}%{doi=false,isbn=false,url=false}
\bibliography{sample}

% Full bibliography added automatically for Optics Letters submissions
% Note that this extra page will not count against page length
%\ifthenelse{\equal{\journalref}{ol}}{%
\clearpage
%\bibliographyfullrefs{sample}

% Full bibliography added automatically for Optics Letters submissions
% Note that this extra page will not count against page length
%\ifthenelse{\equal{\journalref}{ol}}{
%\clearpage
%\bibliographyfullrefs{sample}
%}{}
 
%Manual citation list
%\begin{thebibliography}{1}
%\bibitem{Zhang:14}
%Y.~Zhang, S.~Qiao, L.~Sun, Q.~W. Shi, W.~Huang, %L.~Li, and Z.~Yang,
 % \enquote{Photoinduced active terahertz metamaterials with nanostructured
  %vanadium dioxide film deposited by sol-gel method,} Opt. Express \textbf{22},
  %11070--11078 (2014).
%\end{thebibliography}

\end{document}